\documentclass[twocolumn,showpacs,preprintnumbers,amsmath,amssymb]{revtex4}

\usepackage{graphicx}
\usepackage{dcolumn}
\usepackage{bm}

\begin{document}

\preprint{APS/123-QED}

\title{Systematic study of multi-fragmentation in asymmetric colliding nuclei  
\\}

\author{Varinderjit Kaur}
\author{Suneel Kumar}%
\email{suneel.kumar@thapar.edu}

\affiliation{%
School of Physics and Material Science, Thapar University Patiala-147004, Punjab (India)\\
}

\date{\today}
\begin{abstract}
We present a complete systematically theoretical study of multifragmentation for asymmetric colliding 
nuclei for heavy-ion reactions in the energy range between 
50 MeV/nucleon and 600 MeV/nucleon by using soft and hard equations of state. This study is performed 
within an isospin-dependent quantum molecular dynamics model.
To see the effect of mass asymmetry, simulations are carried out in the absence of 
Coulomb interactions. Coulomb interactions enhances the production of fragments by about 20${\%}$.
We envision an interesting outcome for large asymmetric colliding nuclei. Although nearly symmetric nuclei 
depict a well known trend for rising and falling with a peak around E= 100 MeV/nucleon, 
this trend, however, is completely missing for large asymmetric nuclei. Therefore, experiments are needed 
to verify this prediction.  
\end{abstract}
\maketitle
\baselineskip=1.5\baselineskip
\section{Introduction}
The ability to understand the properties of nuclear matter at the extreme conditions of temperature and 
density is one of the challenges in present-day nuclear-physics research. Such extreme 
conditions can be generated in
a heavy-ion-induced reaction at intermediate energies \cite{reports,hartnack}. The outcome 
of a reaction depends on 
the incident energy, the impact parameter, as well as on the asymmetry of the colliding partners 
\cite{reports,hartnack,ogilivie,PRC58,PRC54}. For symmetrically heavy colliding nuclei at central 
impact parameters, two primary fragments are formed: one that is the projectilelike fragment and the
other that is the targetlike fragment. These excited fragments deexcite through various exit channels: 
evaporation of light particles and emission of intermediate- mass fragments (IMFs) 
\cite{reports,hartnack,ogilivie,PRC58,PRC54}. The excitation energy deposited in the system
at low incident energies is too small to allow the break up of the nuclei into fragments. With an 
increase in the incident energy, colliding nuclei may break into dozens of fragments consisting 
of light, medium, and heavy fragments. The size of the fragments and physics behind their formation 
differs in different physical conditions. No such fragments will survive at extremely high incident
energies. \\ 
Among various theoretical models developed to study these reactions, one can group them into those, 
which are statistical in nature \cite{bondorf}, and others, which take the dynamics of the reaction into 
account and, hence, are capable of investigating the evolution of the fragmentation
and nucleon-nucleon correlations \cite{reports,hartnack,ogilivie,PRC58,PRC54,Li,fuchs}. Interestingly, 
both types of models (although different in their assumptions) are able to explain one or the other 
feature of the experimental 
findings. Here, we will only concentrate on the dynamical model. A careful analysis of experimental
efforts reveals either that one has studied the collision of symmetric nuclei 
(e.g., $_{79}Au^{197}+_{79}Au^{197}$) \cite{ogilivie} or that one has studied
asymmetric colliding nuclei(e.g., $_{20}Ar^{40}+_{21}Sc^{45}$, $_{20}Ar^{40}+_{79}Au^{197}$
$_{6}C^{12}+_{79}Au^{197}$, $_{20}Ca^{40}+_{79}Au^{197}$) 
\cite{ogilivie}. Although the dynamics for symmetrically heavy nuclei is prominently exposed in
experimental and theoretical studies, little attention is paid to the collision of asymmetric nuclei. 
We, at the same time, know that the 
symmetry and the isospin play decisive roles in a reaction. We want to discover the fragmentation of
asymmetrically colliding pairs in the following different ways.  
(i) In the first case, we will perform a systematic study of the emission of various fragments
as a function of the asymmetry ${\eta}$ of a reaction. The asymmetry of a reaction can be defined by the 
asymmetry parameter $\eta = {(A_T-A_P)}/{(A_T+A_P)}$ \cite{rkpuri,gupta}; ${A_T}$ and ${A_P}$ are, 
respectively, 
the masses of the target and the projectile. ${\eta = 0}$ corresponds to the symmetric reaction, whereas
nonzero values of ${\eta}$ define different asymmetries of the reaction. It is worth mentioning 
that the outcome and the physical mechanism behind the symmetric and asymmetric reactions are 
entirely different \cite{ogilivie,rkpuri,gupta}. Here, for systematic analysis, we start from the 
symmetrically colliding partners
(${\eta = 0}$), and then, asymmetry parameter ${\eta}$ is varied gradually (${\eta}$ = -0.8 to 0.8) 
by keeping the total mass of the system fixed. Such an experiment was performed by Betts 
\cite{betts} in 1981, where fusion probabilities were measured for different colliding 
pairs, which lead to the same compound nucleus.\\  
(ii) In the second case, the projectile mass is varied from 16 to 56 units, while the total mass of 
the system is kept fixed. For example, we study the reactions of $_{8}O^{16}+_{54}Xe^{136}$, 
$_{14}Si^{28}+_{54}Xe^{124}$, $_{16}S^{32}+_{50}Sn^{120}$, 
$_{20}Ca^{40}+_{50}Sn^{112}$, $_{24}Cr^{50}+_{44}Ru^{102}$, $_{26}Fe^{56}+_{44}Ru^{96}$, etc. The target
isotope is chosen to be a stable one. Since, we will neglect the Coulomb effect, we may say that it leads 
to the same compound nucleus. \\
In a recent communication 
\cite{jian}, Liu studied the isospin effects on the process of multi-fragmentation and dissipation 
by considering the two pairs of colliding systems ${Zn^{76}+Ar^{40}}$ and ${Kr^{76}+Ca^{40}}$, 
${Cd^{120}+Ar^{40}}$ and ${Xe^{120}+Ca^{40}}$ for central collisions. Another study \cite{jian} 
focused on the isospin effects of the mean-field and two-body collisions on nucleon emissions 
at intermediate energies. This study showed that the neutron-proton ratio of preequilibrium 
nucleon emission and the neutron-proton differential and elliptical flows are the probes for 
extracting the isospin-dependent mean field at a lower beam-energy region. Because of less compression 
in asymmetric reactions, most of the deposited excitation energy is in the form of thermal energy.  
Our present study will shed some light on the effect of the asymmetry of a 
reaction on fragmentation, where a great amount of energy is in the form of thermal energy.\\ 
The present analysis will be carried out within the frame work of the isospin-dependent quantum-molecular 
dynamics (IQMD) model \cite{hartnack,PRC014611}. Our paper is organized as follows:  
We briefly discuss the model in Sec.II. Our results are 
given in Sec.III, and we summarize the results in Sec.IV.
\section{The Model}

The IQMD \cite{hartnack} model treats different 
charge states of nucleons, deltas, and pions explicitly \cite{HARTNACK1}, as shown in the 
Vlasov-Uehling-Uhlenbeck (VUU) model 
\cite{KRUSE}. The IQMD model was successfully used in analyzing the large number of observables 
from low to relativistic energies \cite{hartnack,PRC58,PRC54,PRC014611,
HARTNACK1,KRUSE}. One of its versions (quantum-molecular dynamics) has been very successful in explaining 
the subthreshold particle 
production \cite{huang}, the multifragmentation \cite{PRC58,fuchs}, the collective flow \cite{PRC58,sood}, 
the disappearance of flow \cite{sood}, and the density temperature reached in a reaction \cite{fuchs}. 
We will not take relativistic effects into account, since there is no effect \cite{lehmann} in the energy
domain in which we are interested. The isospin degree 
of freedom enters into the calculations via both cross sections and mean field  
\cite{KRUSE}. The details about the elastic and inelastic cross sections for 
proton-proton and neutron-neutron collisions can be found in Refs.\cite{hartnack,lehmann}. \\
In this model, baryons are represented by Gaussian-shaped density distributions \\
\begin{equation}
f_i(r,p,t) = \frac{1}{{\pi}^2{\hbar}^2}e^{\frac{{-(r-r_i(t))^2}}{2L}}e^{\frac{{-(p-p_i(t))^2}.2L}{\hbar^2}}.
\end{equation}
Nucleons are initialized in a sphere with radius R = ${1.12A^{1/3}}$ fm, in accordance with the  liquid- 
drop model. Each nucleon occupies a volume of ${\hbar^3}$ so that phase space is uniformly filled.
The initial momenta are randomly chosen between 0 and Fermi momentum ${p_F}$. The nucleons of the target 
and the projectile interact via two- and three-body Skyrme forces and the Yukawa potential. The isospin 
degrees of freedom are treated explicitly by employing a symmetry potential and the explicit Coulomb forces between 
protons of the colliding target and protons of the projectile. This helps to achieve the correct
distribution of protons and neutrons within the nucleus.\\
The hadrons propagate by using Hamiltonian equations of motion:\\
\begin{equation}
\frac{d\vec{r_i}}{dt} = \frac{d<H>}{d\vec{p_i}}~~~~;~~~~\frac{d\vec{p_i}}{dt} = -\frac{d<H>}{d\vec{r_i}}.
\end{equation}
 with\\ 
${ <H> = <T> + <V>}$  as the Hamiltonian.
\begin{eqnarray}
    =  \sum_i\frac{p_i^2}{2m_i} + \sum_i \sum_{j>i}\int f_i(\vec{r},\vec{p},t)V^{ij}(\vec{r'},\vec{r})\nonumber\\ 
f_j(\vec{r'},\vec{p'},t)d\vec{r}d\vec{r'}d\vec{p}d\vec{p'}.
\end{eqnarray}
The baryon-baryon potential ${V^{ij}}$, in the preceding relation, reads as\\
\begin{eqnarray}
V^{ij}(\vec{r'}-\vec{r}) &~=~& V_{Skyrme}^{ij} + V_{Yukawa}^{ij} + V_{Coul}^{ij} + V_{Sym}^{ij}\nonumber\\
&=&t_1\delta(\vec{r'}-\vec{r})+ t_2\delta(\vec{r'}-\vec{r}){\rho}^{\gamma-1}(\frac{\vec{r'}+\vec{r}}{2})\nonumber\\ 
&+& t_3\frac{exp({\mid{\vec{r'}-\vec{r}}\mid}/{\mu})}{({\mid{\vec{r'}-\vec{r}}\mid}/{\mu})} + \frac{Z_{i}Z_{j}{e^2}}{\mid{\vec{r'}-\vec{r}\mid}}\nonumber\\
&+& t_{4} \frac{1}{\rho_o}T_{3}^{i}T_{3}^{j}.\delta(\vec{r'_i} - \vec{r_j}).
\end{eqnarray}
Where ${{\mu} = 0.4 fm}$, ${t_3 = -6.66 MeV}$, and ${t_4 = 100 MeV}$. The values of ${t_1}$ and ${t_2}$ 
depends on the values of ${\alpha}$, ${\beta}$, and ${\gamma}$ \cite{reports}.
Here, ${Z_i}$ and ${Z_j}$ denote the charges of the ${i^{th}}$ and ${j^{th}}$ baryons, and ${T_{3}^i}$, 
${T_{3}^j}$ are their respective ${T_3}$ components (i.e. 1/2 for protons and -1/2 for neutrons). 
The Meson potential only consists of the Coulomb interaction. The parameters ${\mu}$ and ${t_1,........,t_4}$ 
are adjusted to the real part of
the nucleonic optical potential. For the density dependence of the nucleon optical potential, standard 
Skyrme-type parametrizations are employed. The Skyrme energy density has been shown to be very
successful at low incident energies, where fusion is the dominant channel \cite{rkpuri,gupta}. 
The Yukawa term is quite
similar to the surface-energy coefficient used in the calculations of the nuclear potential for fusion
\cite{ishwar}. 
The choice of the equations of state (EOS) (or compressibility) is still a 
controversial one. Many studies advocate softer matter, whereas, a greater number of studies indicate matter to be 
harder in nature \cite{KRUSE, sood}. We will use both hard (H) and soft (S) EOS that have compressibilities of 380 and 200 MeV, respectively. \\
The symmetry energy is taken into account by introducing\\ 
\begin{equation}
V_{sym}^{ij} = t_{4} \frac{1}{\rho_o}T_{3}^{i}T_{3}^{j}.\delta(\vec{r'_i} - \vec{r_j}).
\end{equation} 
The binary nucleon-nucleon collisions are included by employing the collision term of the well-known VUU-
Boltzmann-Uehling-Uhlenbeck  
equation \cite{KRUSE, BUU}. The binary collisions are allowed stochastically, in a 
similar manner as performed in 
all transport models. During the propagation, two nucleons are supposed to suffer a binary collision if 
the distance between their centroids is,\\
\begin{equation}
 {\mid{r_i}-{r_j}\mid} \le \sqrt{\frac{\sigma_{tot}}{\pi}},~~~~~~~ {\sigma_{tot}} = \sigma(\sqrt{s}, type),
\end{equation}
where ``type'' denotes the in-going collision partners (N-N, N-${\delta}$, N-${\pi}$...). In addition, Pauli 
blocking (of the final state) of baryons is taken into account by checking the phase space densities 
in the final states. The final phase space fractions ${P_1}$ and ${P_2}$ which are already occupied by 
other nucleons,
are determined for each of the scattering baryons. The collision is then blocked with probability\\
\begin{equation}
{P_{block} = {1- (1-P_1)(1-P_2)}}.
\end{equation}  
The delta decays are checked in an analogous fashion with respect to the phase space of the 
resulting nucleons.\\
\section{Results and Discussions}
In the present calculations, 
a simple spatial clusterization algorithm dubbed as the minimum-spanning-tree 
method is used to clusterize the phase space \cite{reports}. We, however, also acknowledge that more 
microscopic algorithmic routines are also available in literature \cite{hartnack,PRC54}.
By using the asymmetric (colliding) nuclei, the effect of mass asymmetry can be analyzed without 
varying the total mass of the system. As noted previously, the experimental studies by the Michigan 
State University, miniball and 
ALADIN \cite{ALADIN} 
groups vary the asymmetry of the reaction, whereas the plastic ball and FOPI 
experiments \cite{FOPI} are only performed for symmetric 
reactions. \\
The effect of mass asymmetry on fragmentation is demonstrated in Fig.1. Here, relative 
multiplicity ${R_M}$ is defined as = $\mid{\frac{(M_A - M_S)}{M_S}\mid}$ where ${M_A}$ and ${M_S}$ 
are the multiplicities of various fragments obtained in the asymmetric and 
symmetric colliding nuclei, respectively. 
The relative multiplicity of free nucleons, light mass fragments (LMF's) 
${(2\le A \le4)}$, medium mass fragments (MMF's) ${(3\le A \le8)}$, and IMFs ${(5\le A \le{A_{tot}}/6)}$ 
follows hyperbolic behavior. Here, the total mass was kept 
fixed at 140 
\begin{figure}
\includegraphics{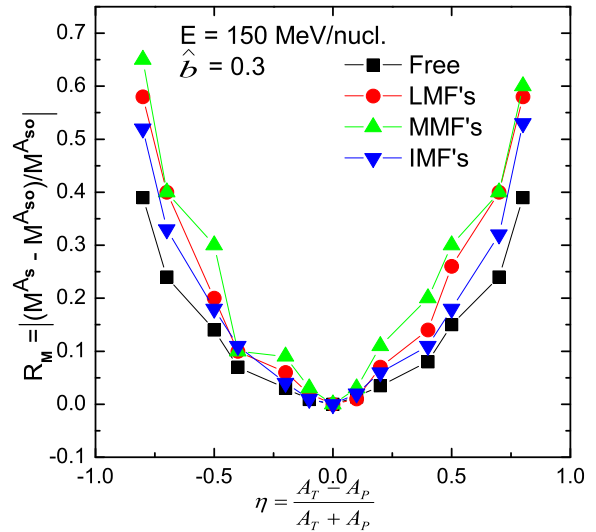}
\caption{\label{fig:1}(Color online) The relative multiplicity of different fragments as a 
function of mass asymmetry ${\eta}$ at E = 150 MeV/nucleon and impact parameter at $\hat{b}$ = 0.3.}  
\end{figure}
units, and the mass of the 
target and the projectile was varied in steps of 10 units (e.g., ${A_P = 130}$, ${A_T = 10}$, ${A_P = 120}$, 
${A_T = 20}$ etc). As asymmetry ${\eta}$ shifts toward the positive side, the target fragmentation 
takes place since  ${A_T >> A_P}$. In contrast, at ${\eta}$ = -0.8, the projectile fragmentation 
takes place since ${A_P >> A_T}$. Since the emission of the nucleons (protons and neutrons)
is maximum in the participant zone, one sees lesser nucleon emission compared to the emission
of lighter fragments such as LMFs and MMFs. In this study, all reactions
are performed in the laboratory frame. Note that the mirror reactions are also studied by the FOPI collaboration 
\cite{mirror}. \\
Now, we confine our study to particular asymmetric systems such as $_{8}O^{16}+_{54}Xe^{136}$, 
$_{14}Si^{28}+_{54}Xe^{124}$, $_{16}S^{32}+_{50}Sn^{120}$, $_{20}Ca^{40}+_{50}Sn^{112}$, 
$_{24}Cr^{50}+_{44}Ru^{102}$, and $_{26}Fe^{56}+_{44}Ru^{96}$ at different incident energies. 
To see the effect of the Coulomb interactions, in Fig.2, we display the final phase 
space of a single event of  $_{26}Fe^{56}+_{44}Ru^{96}$ 
(${\eta}$= 0.2) (upper panel) and $_{14}Si^{28}+_{54}Xe^{124}$ (${\eta}$=0.6) (lower panel) 
at a fixed center-of-mass energy of 250 MeV/nucleon with and without the Coulomb interaction. Here, 
the phase space
of LMFs [${(2\le A \le4)}$], and IMFs
[${(5\le A \le{A_{tot}}/6)}$] is displayed. Irrespective of the Coulomb interaction, the reaction with 
${\eta}$ = 0.2 leads to isotropic emission compared to the reaction with 
 ${\eta}$ = 0.6 that projects a nearly binary character. Since LMFs originate from the midrapidity region, 
they are better suited for studying the effect of asymmetry on the reaction dynamics.\\
Generally, obvious effects associated with the asymmetry of a reaction are caused by thr Coulomb 
interaction. To understand
the role of asymmetry beyond the Coulomb effects, we switch off the Coulomb force in further analysis. 
Additionaly, we keep the center-of-mass energy fixed throughout the analysis. \\
In Fig.3, we compare the effect of the Coulomb forces on the multiplicities
of various fragments at ${E_{c.m.}}$= 50 MeV/nucleon and 
${E_{c.m.}}$= 250 MeV/nucleon. The asymmetry of the reaction is varied by using projectiles with
masses between 16 and 56. We see a clear effect of Coulomb forces. As expected, it is maximum for 
the low incident energies, and this effect diminishes as 
we move toward higher incident energies. An enhanced effect emerges at larger asymmetric reactions. 
\begin{figure}
\includegraphics{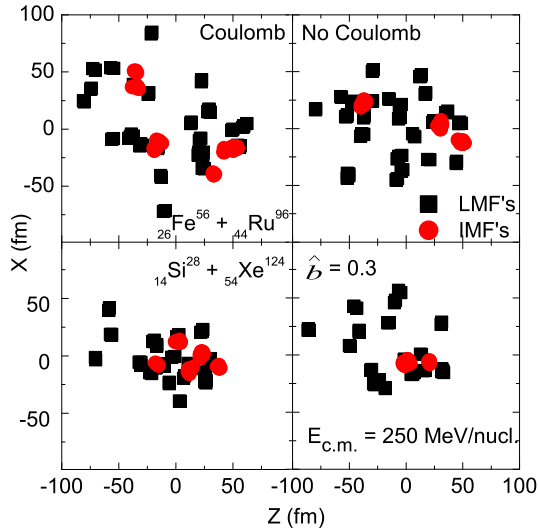}
\caption{\label{fig:2}(Color online) The final-state phase space of a single event for the reaction of 
$_{26}Fe^{56}+_{44}Ru^{96}$ with and without the Coulomb effects. Here, center-of-mass energy is
${E_{c.m.}}$ = 250 MeV/nucleon
 and impact parameter is $\hat{b}$ = 0.3. Different symbols denote LMFs and IMFs.}  
\end{figure}

\begin{figure}
\includegraphics{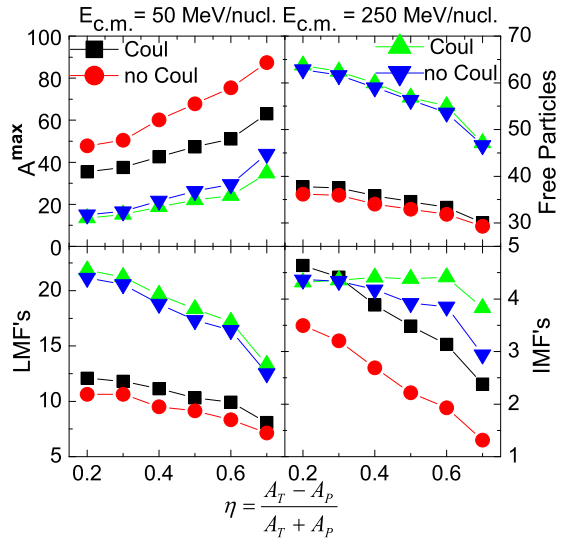}
\caption{\label{fig:3}(Color online) The mass asymmetry ${\eta}$ variation of the production of ${A_{max}}$, 
free nucleons, LMF's and IMF's with and without the Coulomb effect at two different energies at
${E_{c.m.}}$ = 50 MeV/nucleon and ${E_{c.m.}}$ = 250 MeV/nucleon and impact parameter at $\hat{b}$ = 0.3.} 
\end{figure}

The production of the heaviest fragment ${A_{max}}$, the free nucleons, the LMFs 
${(2\le A \le4)}$, and the IMFs ${(5\le A \le{A_{tot}}/6)}$ shows 
expected behavior. The heavier mass continues to grow, and it is close to the mass of the reacting 
partners for larger asymmetries. In contrast, the production of free nucleons, LMFs, 
and IMFs shows a reverse trend with 
the asymmetry of the reaction. This happens because of a decrease in the participant zone. Although
the role of the Coulomb interaction decreases with energy, its effect, however, remains constant  
(~20${\%}$) with the asymmetry of the reaction. Because of the presence of the Coulomb forces, 
the nuclear matter breaks into smaller pieces/free nucleons. \\
To understand this aspect further, we display in Fig.4, the saturation density of the reaction
obtained at 200 fm/c. This density remains quite high for larger asymmetries. This is caused by the
\begin{figure}
\includegraphics{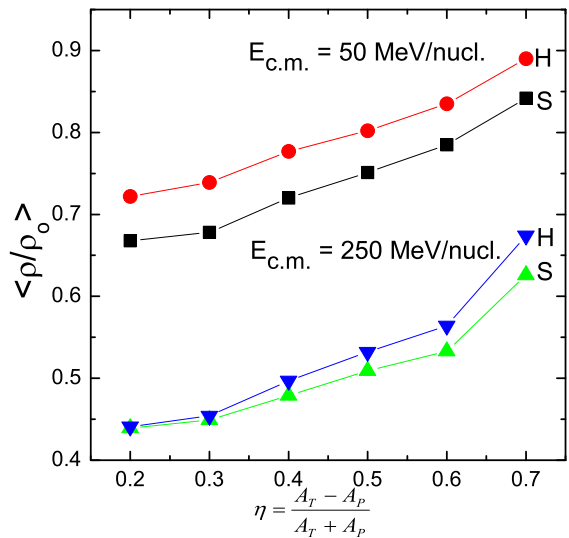}
\caption{\label{fig:4}(Color online) The variation of density with mass asymmetry at an impact parameter 
$\hat{b}$ = 0.3. The different symbols are at ${E_{c.m.}}$ = 50 MeV/nucleon and ${E_{c.m.}}$ 
= 250 MeV/nucleon with S and H EOS.}  
\end{figure}
least amount of destruction of nuclear matter at larger asymmetries, which leads to higher
nucleonic densities. As a result, one sees a heavier ${A_{max}}$ compared to the smaller ${\eta}$ values.
Because of the heavier ${A_{max}}$, the emission of the nucleons is also smaller.\\

In Fig.5, the charge distribution is displayed as a function of the fragment charge using S 
(upper panel) and H EOS (lower panel). The symmetric reactions (${\eta}$=0) lead to
enhanced emission of nucleons and LMFs compared to asymmetric reactions, where incomplete fusion 
events or deep inelastic events are dominant. We can also say that a large asymmetric reaction leads to
few nucleonic-transfer processes. One also notices that the slope of the distribution 
becomes steeper with H EOS compared to S EOS. Because of the enhanced binary collisions
between the nucleons for nearly symmetric nuclei, the emission of fragments is suppressed.
Clear systematics can be seen in the production of fragments with the asymmetry of the reaction. \\
\begin{figure}
\includegraphics{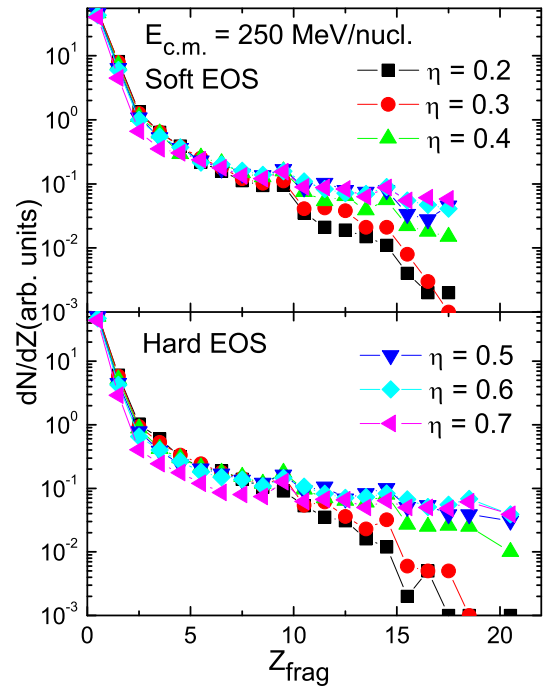}
\caption{\label{fig:5}(Color online) The charge distribution for different asymmetries between ${\eta}$ = 
0.2 to 0.7 at ${E_{c.m.}}$ = 150 MeV/nucleon and impact parameter at $\hat{b}$ = 0.3. 
The upper and lower panels are shown with S and H EOS, respectively.}  
\end{figure}
In Fig.6, the variation of the multiplicity of LMFs is displayed as a function 
of the center-of-mass energy for various asymmetric reactions using H and S EOS. Because of
more compression, the nearly symmetric reaction drives matter into the participant zone and, as a result,
more lighter fragments are emitted.\\       
\begin{figure}
\includegraphics{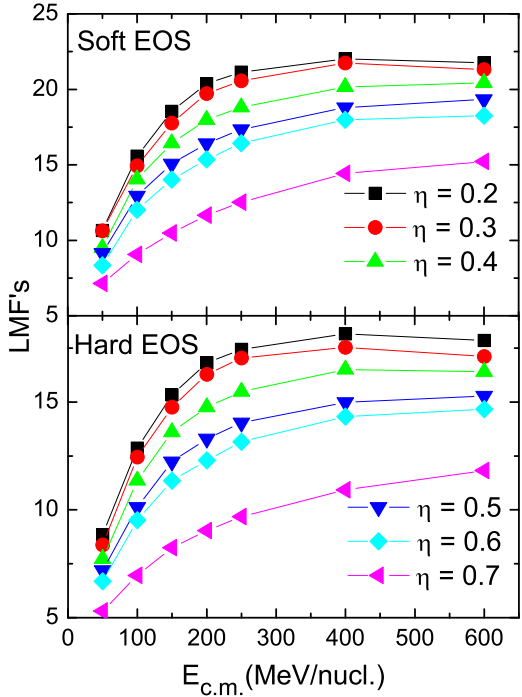}
\caption{\label{fig:6}(Color online) The variation of LMFs with center-of-mass 
energy using S and H EOS at impact parameter $\hat{b}$ = 0.3. 
The different symbols show the results, which involve different asymmetries. }  
\end{figure}

In Fig.7, the variation of the multiplicity of IMFs is displayed as a 
function of the center-of-mass energy ${E_{c.m.}}$. 
\begin{figure}
\includegraphics{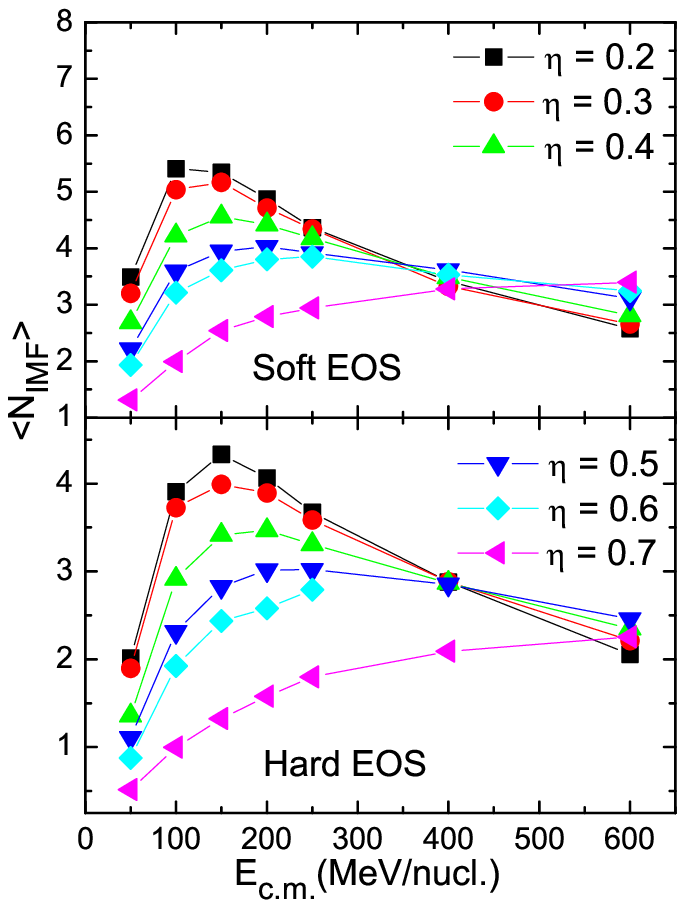}
\caption{\label{fig:7}(Color online) Same as Fig.6, but for IMFs. }  
\end{figure}
This happens because of the fact that the system suffers less compression; and, hence, less numbers of 
IMFs are produced. One notices several interesting points: The nearly
symmetric collision leads to a well-known trend (i.e., the maximum emission occurs around
100 MeV/nucleon). This trend, however is not shown by the large asymmetric reactions where
we donot see any sharp rise or fall; and, furthermore, a flat plateau is obtained at much higher incident
energies compared to nearly symmetric nuclei. Therefore, experiments are needed to verify this prediction.  \\

\section {Conclusion}
A systematically theoretical study is presented for the asymmetric colliding nuclei, which use a variety 
of reactions that employ different EOS as well as incident energies. We envision an interesting outcome for
large asymmetric colliding nuclei, although nearly symmetric nuclei depict a well-known trend of rising 
and falling with peak around E= 100 MeV/nucleon. This trend, however, is completely missing for large
asymmetric nuclei. In conclusion, experiments are needed to verify this prediction.\\

\section {Acknowledgment}
This work has been supported by the grant from Department of Science and Technology (DST) Government of 
India vide Grant No.SR/WOS-A/PS-10/2008.\\ 


\section{References}
\begin{thebibliography}{999}
\bibitem{reports} J. Aichelin and H. Stocker, Phys. Lett. B {\bf 176}, 14 (1986); 
J. Aichelin, Phys. Report {\bf 202}, 233 (1991); 
W. Cassing, W. Metag, U. Mosel and K. Nitta, Phys. Rep. {\bf 188}, 363(1990);
G. F. Bertsch and S. Das. Gupta, Phys. Rep. {\bf 160}, 189 (1988);
H. Feldmeier, Nucl. Phys. {\bf 515}, 147 (1990);
J. Schnack, Ph. D. Thesis, GSI report, Darmstadt (1996);
A. Ono, H. Horiuchi, T. Maruyama, Phys. Rev. C {\bf 48}, 2946 (1993);
{\it ibid.}  {\bf 47}, 2652 (1993);
S. A. Bass, Ch. Hartnack, H. Stocker, and W. Greiner, Phys. Rev. C {\bf 51}, 3343 (1995); 
B. J. VerWest and R. A. Arndt, Phys. Rev. C {\bf 25}, 1979 (1982);
B. Jakobsson {\it et al.}, Nucl. Phys. A {\bf 509}, 195 (1990); 
H. W. Barz {\it et al.}, {\it ibid.} {\bf 548}, 427 (1992).

\bibitem{hartnack} Ch. Hartnack, R. K. Puri, J. Aichelin, J. Konopka, S. A. Bass, H. Stocker, and
W. Greiner, Eur. Phys. J. A {\bf 1}, 151 (1998);
Li. Zhuxia, Ch. Hartnack, H. Stocker and W. Greiner, Phys. Rev. C {\bf 44}, 824 (1991);
P. B. Gossiaux, R. K. Puri, Ch. Hartnack, and J. Aichelin, Nucl. Phys. A {\bf 619}, 379 (1997);
S. Kumar and R. K. Puri, Phys. Rev. C {\bf 58}, 320 (1998);
{\it ibid.} {\bf 58}, 2858 (1998);
{\it ibid.} {\bf 60}, 054607 (1999).
\bibitem{ogilivie} C. A. Ogilivie {\it et al.}, Phys. Rev. Lett. {\bf 67}, 1214 (1991); 
M. B. Tsang {\it et al.}, {\it ibid.} {\bf 71}, 1502 (1993); 
R. T. de Souza {\it et al.}, Phys. Lett. B {\bf 268}, 6 (1991); 
C. Williams {\it et al.}, Phys. Rev. C {\bf 55}, R2132(1997); 
N. T. B. Stone {\it et al.}, Phys. Rev. Lett. {\bf 78}, 2084 (1997);  
J. Hubble {\it et al.}, Z. Phys. A {\bf 340}, 263 (1991);
W. J. Llope {\it et al.}, Phys. Rev. C {\bf 51}, 1325 (1995);
Th. Rubehn {\it et al.}, Phys. Rev. C {\bf 53}, 993 (1996).
\bibitem{PRC58} S. Kumar, M. K. Sharma, and R. K. Puri, Phys. Rev. C {\bf 58}, 3494 (1998); 
S. Kumar, R. K. Puri, and J. Aichelin {\it ibid.} {\bf 58}, 1618 (1998);
S. Kumar, S. Kumar, and R. K. Puri {\it ibid.} {\bf 78}, 064602 (2008); 
{\it ibid.} {\bf 81}, 014601 (2010);
J. K. Dhawan and R. K. Puri, {\it ibid.} {\bf 75}, 057601 (2007);
{\it ibid.} 057901 (2007).
\bibitem{PRC54} R. K. Puri, Ch. Hartnack and J. Aichelin, Phys. Rev. C {\bf 54}, R28 (1996);
R. K. Puri and J. Aichelin, J. Comput. Phys. {\bf 162}, 245 (2000);
Y. K. Vermani, J. K. Dhawan, S. Goyal, R. K. Puri, and J. Aichelin, J. Phys. G: Nucl. Part. Phys. {\bf 37}, 015105 (2010);
Y. K. Vermani and R. K. Puri, Europhys. Lett. {\bf 85}, 62001 (2009);
Y. K. Vermani, S. Goyal, and R. K. Puri, Phys. Rev. C {\bf 79}, 064613 (2009). 

\bibitem{bondorf} J. P. Bondorf {\it et al.}, Nucl. Phys. A {\bf 443}, 321 (1985); 
D. H. E. Gross , Phys. Prog. Rep. {\bf 53}, 605 (1990); 
A. Botvina {\it et al.}, Z. Phys. A {\bf 345}, 297 (1993).
\bibitem{Li} B. A. Li and D. H. E. Gross, Nucl. Phys. A {\bf 554}, 257 (1993); 
H. M. Xu, C. A. Gagliardi, R. E. Tribble and C. Y. Wong, {\it ibid.} {\bf 569}, 575 (1994); 
F. Daffin, K. Haglin and W. Bauer, Phys. Rev. C {\bf 54}, 1375 (1996); 
V. de la Mota, F. Sebille, M. Farine, B. Remaud and P. Schuck, {\it ibid.} {\bf 46}, 677 (1992); 
\bibitem{fuchs} C. Fuchs, E. Lehmann, R. K. Puri, L. Sehn, A. Faessler, and H. H. Wolter, 
J. Phys. G: Nucl. Part. Phys. {\bf 22}, 131 (1996);
Y. K. Vermani and R. K. Puri, {\it ibid.} {\bf 36}, 105103 (2009); 
R. K. Puri, N. Ohtsuka, E. Lehmann, A. Faessler, M. A. Matin, D. T. Khoa, G. Batko, and S. W. Huang,
Nucl. Phys. A {\bf 575}, 733 (1994);
A. D. Sood and R. K. Puri, Phys. Rev C {\bf 79}, 064618 (2009).
\bibitem{rkpuri} R. K. Puri and N. Dhiman, Eur. Phys. J. A {\bf 23}, 429 (2005);
R. Arora, R. K. Puri, and R. K Gupta, {\it ibid.} {\bf 8}, 103 (2000);
R. K. Puri, M. K. Sharma, and R. K. Gupta, {\it ibid.} {\bf 3}, 277 (1998);
R. K. Puri and R. K. Gupta, Phys. Rev. C {\bf 51}, 1568 (1995);
R. K. Puri, P. Chattopadhyay, and R. K. Gupta {\it ibid.} {\bf 45}, 1837 (1992);
{\it ibid.} {\bf 43}, 315 (1991);
R. K. Puri and R. K. Gupta, J. Phys. G: Nucl. Part. Phys. {\bf 18}, 903 (1992).
\bibitem{gupta} R. K. Gupta, S. Singh, R. K. Puri, and W. Scheid, Phys. Rev. C {\bf 47}, 561 (1993);
 R. K. Gupta, S. Singh, R. K. Puri, A. Sandulescu, W. Greiner and W. Scheid, 
J. Phys. G {\bf 18}, 1533 (1992);
S. S Malik, S. Singh, R. K. Puri, S. Kumar and R. K. Gupta, Pramana J. Phys. {\bf 32}, 419 (1989);
R. K. Puri, S. S. Malik, and R. K. Gupta, Europhys. Lett. {\bf 9}, 767 (1989).
\bibitem{betts} R. R. Betts, Proc. Conf. on Resonances in Heavy Ion Reactions (Lecture Notes in Physics
156, 1981) {\it ed.} K. A. Eberhardt (Berlin: Springer) p 185.
\bibitem{jian} J. Y. Liu, Y. F. Yang, W. Zho, S. W. Wang, Q. Zhao, W. J. Guo, and B. Chen, 
Phys. Rev. C {\bf 63}, 054612 (2001); 
J. Y. Liu, Y. Z. Xing, and W. J. Guo, Chin. Phys. Lett. {\bf 20}, 643 (2003).
\bibitem{PRC014611} S. Kumar, S. Kumar, and R. K. Puri, Phys. Rev. C {\bf 81}, 014611 (2010).
\bibitem{HARTNACK1} C. Hartnack, H. Oeschler, and J. Aichelin, Phys. Rev. Lett. {\bf 90}, 102302 (2003);
{\it ibid.} J. Phys. G {\bf 35}, 044021 (2008).
\bibitem{KRUSE} H. Kruse, B. V. Jacak, H. Stocker, Phys. Rev. Lett. {\bf 54}, 289 (1985);
J. J. Molitoris and H. Stocker, Phys. Rev. C {\bf 32}, R346 (1985);
J. Aichelin and G. Bertsch, Phys. Rev. C {\bf 31}, 1730 (1985);
Ch. Hartnack, H. Oeschler, and J. Aichelin, Phys. Rev. Lett. {\bf 96}, 012302 (2006);
A. D. Sood and R. K. Puri, Phys. Rev. C {\bf 69}, 054612 (2004);
A. D. Sood and R. K. Puri, Phys. Rev. C {\bf 73}, 067602 (2006);
G. D. Westfall {\it et al.}, Phys. Rev. Lett. {\bf 71}, 1986 (1993);
D. J. Magestro, W. Bauer, and G. D. Westfall, Phys. Rev. C {\bf 62}, 041603(R) (2000). 
\bibitem{huang} S. W. Huang {\it et al.}, Prog. Part. Nucl. Phys. {\bf 30}, 105 (1993);
G. Batko {\it et al.}, J. Phys. G: Nucl. Part. Phys. {\bf 20}, 461 (1994).
\bibitem{sood} A. D. Sood and R. K. Puri, Phys. Rev. C {\bf 70}, 034611 (2004);
{\it ibid.} Phys. Lett. B {\bf 594}, 260 (2004);
E. Lehmann, A. Faessler, J. Zipprich, R. K. Puri, and S. W. Huang, Z. Phys. A {\bf 355}, 55 (1996).

\bibitem{lehmann} E. Lehmann, R. K. Puri, A. Faessler, G. Batko, and S. W. Huang, 
Phys. Rev. C {\bf 51}, 2113 (1995);
{\it ibid.} Prog. Part. Nucl. Phys. {\bf 30}, 219 (1993).
\bibitem{ishwar} I. Dutt and R. K. Puri Phys. Rev. C {\bf 81}, 047601 (2010); {\it ibid.} 
{\bf 81}, 044615 (2010).
\bibitem{BUU} H. Stocker and W. Greiner, Phys. Rep. {\bf 137}, 277 (1986);
A. D. Sood and R. K. Puri, Phys. Rev. C {\bf 69}, 054612 (2004);
D. J. Magestro, W. Bauer, O. Bjarki, J. D. Crispin, M. L. Miller, M. B. Tonjes, A. M. VanderMolen, 
G. D. Westfall, R. Pak , and E. Norbeck, Phys. Rev. C {\bf 61}, 021602(R) (2000);
P. Danielewicz, R. Lacey, and W. G. Lynch, Science {\bf 298}, 1592 (2002).
\bibitem{ALADIN}  D. R. Bowman {\it et al.}, Phys. Rev. C {\bf 46}, 1834 (1992);
A. Le Fevre {\it et al.}, Nucl. Phys. A {\bf 735}, 219 (2004);
J. Lukasik, {\it et al.}, Phys. Lett. B {\bf 566}, 76 (2003);
K. Turzo {\it et al.}, Eur. Phys. J. A {\bf 21}, 293 (2004);
C. Volant {\it et al.}, Nucl. Phys. A {\bf 734}, 545 (2004); 
A. Sch\"uttauf {\it et al.}, Nucl. Phys. A {\bf 607}, 457 (1996);
J. Hubele {\it et al.}, Phys. Rev. C {\bf 46}, R1577 (1992).

\bibitem{FOPI} X. Lopez {\it et al.}, Phys. Rev. C {\bf 76}, 052203R (2007);
A. Andronic {\it et al.}, Phys. Lett. B {\bf 612}, 173 (2005);
H. H. Gutbrod {\it et al.}, Z. Phys. A {\bf 337}, 57 (1990);
A. Andronic {\it et al.}, Nucl. Phys. A {\bf 679}, 765 (2001);
J. Lukasik {\it et al.}, Phys. Lett. B {\bf 608}, 223 (2005);
J. Lukasik {\it et al.}, (INDRA Collaboration), presented at the International Workshop on 
Multifragmentation and related Topics (IWM2003), Caen, France, 2003 (unpublished).
\bibitem{mirror} C. Roy {\it et al.}, Z. Phys. A {\bf 358}, 73 (1997).
\end{thebibliography}
\end{document}